\def\lesssim{\mathrel{\hbox{\rlap{\hbox{\lower4pt\hbox{$\sim$}}}\hbox{$<$}}}}
\def\gtrsim{\mathrel{\hbox{\rlap{\hbox{\lower4pt\hbox{$\sim$}}}\hbox{$>$}}}}
\def\msun{$M_{\odot}$}

\def\ll_lsun{log$({L/\rm L_{\sun}})$~}
\def\masa_msun{$M/ \rm M_{\sun}$~}
\def\m_mstar{$M/M_{*}$~}

\documentclass[traditabstract]{aa} % for the abstract without structuration 
\usepackage{graphicx}
\usepackage{txfonts}
\usepackage{graphicx}
        
\begin{document}

\title{Modeling He-rich subdwarfs through the hot-flasher scenario.}
\author{M. M. Miller Bertolami$^{1,2}$\thanks{Fellow of CONICET, Argentina.},
        L. G. Althaus$^{1,2}$\thanks{Member of the Carrera del Investigador
        Cient\'{\i}fico y Tecnol\'ogico, CONICET, Argentina},
	K. Unglaub$^3$,
	A. Weiss$^{4}$}
\offprints{M. M. Miller Bertolami}
\institute{Facultad de Ciencias  Astron\'omicas y Geof\'{\i}sicas,
           Universidad Nacional de La Plata, Paseo del Bosque s/n,
           (1900) La Plata, Argentina.\ \and Instituto de
           Astrof\'{\i}sica La Plata, UNLP-CONICET \and
           Dr. Remeis-Stenwarte Bamberg, Sterwartstr. 7, D96049
           Bamberg, Germany \and Max-Planck-Institut f\"ur
           Astrophysik, Karl-Schwarzschild-Str. 1, 85748, Garching,
           Germany.\\
\email{mmiller@fcaglp.unlp.edu.ar}}
\date{\today}
\abstract
{We present 1D numerical simulations aimed at studying the hot-flasher
scenario for the formation of He-rich subdwarf stars.  Sequences were
calculated for a wide range of metallicities and physical assumptions,
such as the stellar mass at the moment of the helium core flash. This
allows us to study the two previously proposed flavors of the
hot-flasher scenario (``deep'' and ``shallow'' mixing cases) and to
identify a third transition type.  Our sequences are calculated by
solving simultaneously the mixing and burning equations within a
diffusive convection picture, and in the context of standard mixing
length theory.  We are able to follow chemical evolution during
deep-mixing events in which hydrogen is burned violently, and
therefore able to present a homogeneous set of abundances for
different metallicities and varieties of hot-flashers.  We extend the
scope of our work by analyzing the effects of non-standard
assumptions, such as the effect of chemical gradients, extra-mixing at
convective boundaries, possible reduction in convective velocities, or
the interplay between difussion and mass loss. Particular emphasis is
placed on the predicted surface properties of the models.

 We find that the hot-flasher scenario is a viable explanation for the
formation and surface properties of He-sdO stars. Our results also
show that, during the early He-core burning stage, element diffusion
may produce the transformation of (post hot-flasher) He-rich
atmospheres into He-deficient ones. If this is so, then we find that
He-sdO stars should be the progenitors of some of the hottest sdB
stars. }
%{TRADITIONAL ABSTRACT} 
%{TRADITIONAL ABSTRACT} 
%{TRADITIONAL ABSTRACT} 
%{TRADITIONAL ABSTRACT}
\keywords{stars:  evolution  ---  Stars: horizontal-branch --- Stars: subdwarfs --- Stars: mass-loss}
\authorrunning{M. M. Miller Bertolami, et al.}
\titlerunning{Modeling He-rich subdwarfs through the hot-flasher scenario.}
\maketitle

%----------------------------------------------------------------
\section{Introduction}
Hot subluminous stars are an important population of faint blue stars
that can roughly be grouped into the cooler sdB stars, whose spectra
display no or only weak helium lines, and the hotter sdO stars, which
have higher helium abundances and can even be dominated by
helium. Subluminous O, B stars were identified with stars
populating the hot end of the horizontal branch (HB) of some globular
and open clusters (de Boer et al. 1997). While the sdB stars form a
homogeneous spectroscopic class, a large variety of spectra is
observed among sdO stars (Lemke et al.  1997). This diversity of
spectra and the helium-enhanced surface abundances observed in many
of these stars pose a challenge to our understanding of stellar evolution. As a
consequence, some non-canonical evolutionary scenarios were
proposed for their formation.  Among them, the merger of two white
dwarfs (Saio \& Jeffery 2000) and a core helium flash after 
departure from the red giant branch (i.e. ``hot-flasher scenario'')
offer the most promising explanations of their formation (Str\"oer
et al. 2007, Moehler et al. 2007, and Hirsch et al. 2008). 
The second possibility is the subject of the present article. The
hot-flasher scenario was proposed to explain the existence and
characteristics of blue hook stars (such as helium or carbon
enhancement) of some globular clusters e.g. $\omega$Cen and NGC 2808
(D'Cruz et al. 2000, Moehler et al. 2002, Moehler et al. 2004, and
Moehler et al. 2007). The coexistence of extreme horizontal branch
(EHB) stars and helium core white dwarfs in globular clusters
(Calamida et al. 2008) also agrees with a natural prediction of the
hot-flasher scenario (see Castellani et al.  2006). Although in this
scenario mass loss must be increased artificially, from values usually
adopted in the modeling of low mass AGB stars, the driving mechanism
behind RGB mass loss is not well understood (Espey \& Crowley
2008). Differences in metallicity, initial He abundances or rotational
velocities may, in principle, produce enhanced mass loss rates (the
last two by an increase in the RGB-tip luminosity). The later is
particularly interesting in view of the fact that some He-sdO stars
are known to be relatively fast rotators (in comparison with sdB
stars; Hirsch et al. 2008).  The existence of high mass-loss rates in
old metal-rich populations is also an important ingredient of one of
the most successful hypothesis for modeling the UV-upturn of
elliptical galaxies (Yi \& Yoon 2004).  If this is so, it is very
probable that a fraction of stars in these populations will undergo
the helium core flash (HeCF) after departing from the first red giant
branch (RGB) --- i.e. will be ``hot-flashers'' as coined by D'Cruz et
al. (1996).  Finally, in high stellar density enviroments, it is
possible that enhanced mass loss might result from star-to-star
encounters.

In general, low mass stars undergo the HeCF at the tip of the RGB.
However, Castellani \& Castellani (1993) showed that, if sufficient
mass is lost on the RGB, the star can depart from the RGB and
experience the HeCF while descending the hot white dwarf cooling
track. Afterwards D'Cruz et al. (1996) analyzed the frequency of these
HeCFs at high temperatures (``Hot Flashers'') by assuming different
values of the mass-loss efficiency and suggested that this scenario
might provide a path for populating the hot end of the HB.  Sweigart
(1997) noticed the similarity between these hot-flashers and the late
helium shell flashes that produce ``born-again AGB stars'' (Iben 1984)
and showed that --- as happens in the born-again scenario --- the
convective zone, generated by the extremely high energy liberation
rate ($L_{\rm He}\sim 10^{10}L_\odot$) during the flash, reaches the
H-rich envelope and engulfs it. As a consequence, much or almost all H
is burned leading to a He-enriched surface abundance. Brown et
al. (2001) explored this scenario to explain anomalies in the hot HB
of NGC 2808. However, it was not until the work of Cassisi et
al. (2003) that simulations including the violent mixing and burning
of H were performed and yields for the abundances of helium, carbon
and nitrogen were calculated consistently.  All of these works were
focused mainly on globular clusters and therefore on low metallicity
progenitors. However, counterparts of these hot-flashers are also
expected in the Galactic disk. This was considered by Lanz et
al. (2004) who assessed the possibility that (field) He-sdB stars
could be formed by the hot-flasher scenario. In this work, the authors
classified three different clases of hot-flashers:(1) early
hot-flashers, which experience the HeCF during the evolution at
constant luminosity in the HR diagram (after departing from the
RGB-tip) and become hot subdwarf stars with standard H/He envelopes;
(2) late hot-flashers with shallow-mixing (SM), which become
He-enriched hot subdwarfs due to convective dilution of the envelope
(at the low-luminosity, low temperature region of the HR diagram) into
deeper regions of the star; and (3) late hot flashers with deep-mixing
(DM) in which the H-rich envelope is engulfed and burned in the
convective zone generated by the primary HeCF.  Due to the numerical
difficulties related to the calculation of the violent H burning 
that occurs during a deep-mixing event (see Cassisi et al. 2003),
Lanz et al. (2004) failed to provide surface abundances in this case.
Finally, Str\"oer et al. (2005, 2007) and Hirsch et al. (2008)
analyzed the possibility that field He-sdO could be generated by this
mechanism and concluded that the idea could not be rejected, although
some serious discrepancies exist between models and
observations. However, the lack of a quantitative study of hot-flasher
events and the chemical composition that arise from them for a wide
range of metallicities, ensures that comparison with observations
remains difficult. We note that only one previous work (Cassisi et
al. 2003) has calculated the abundances produced by the deep-mixing
event but only for one remnant mass and metallicity (Z=0.0015). On the
other hand Lanz et al. (2004) calculated several sequences for
solar-like metallicities but no calculation of the final surface
abundances for the deep-mixing events were presented.  The amount of
observational data for these stars is increasing rapidly as a
byproduct of surveys of faint blue objects such as SDSS and SPY. As
concluded by Cassisi et al. (2003) to assess the viability of the
hot-flasher scenario, a study of the dependence of the hot-flasher
predictions ($T_{\rm eff}$, $g$ and abundances) on the physical
details is needed.

The core feature of this article is to present a homogeneous set of
simulations of the hot-flasher scenario for a wide range of cases in
which the chemical abundances of the models are consistently
calculated.  In particular we present simulations for 4 different
metallicities and different post-RGB remnant masses (which determines
the luminosity of the model at the moment of the HeCF). The work is
organized as follows. In the following section, we discuss the main
ingredients of the simulations performed in this work. We then, in
Sect. 3, discuss different flavors of hot-flasher and He-enrichment
and describe our simulations results under standard assumptions. In
Sect. 4, we analyze deviations from the standard assumptions such as
possible extra-mixing at convective boundaries, the effect of chemical
gradients, or a possible reduction in convective velocities. We also
address the possible effects of element diffusion in the outer
layers. In Sect. 5, we discuss the results of our standard sequences
and compare them with observations. Finally, in Sect. 6, we present
some concluding remarks.

\section{Numerical and physical details}
The sequences presented in this work were calculated with {\tt LPCODE}, a
numerical code for solving the equations of stellar evolution, which
was recently used to model the formation of H-deficient post-AGB stars
in the born-again scenario (Miller Bertolami et al. 2006, and Miller
Bertolami \& Althaus 2007) and, with minor changes, to study global
properties of the evolution of low and intermediate stars (Serenelli
\& Fukugita 2007). {\tt LPCODE} is a Henyey-type stellar evolution code
designed specifically to compute the formation and evolution of white
dwarf stars and described extensively by Althaus et al. (2003, 2005
and references therein). We therefore only briefly refer to the code
to mention some particular features and some modifications of special
interest to this work.

Radiative opacities are those of OPAL (Iglesias \& Rogers 1996),
including opacities for C-rich mixtures as expected for the regions
with violent proton burning during a deep-mixing event (Cassisi et
al. 2003) complemented at low temperatures with the molecular
opacities of Alexander \& Ferguson (1994). In the present work,
conductive opacities for the degenerate He-core were included
according to Cassisi et al. (2007). As a consequence, we note that the
core masses of our models at the HeCF are slightly lower than in
previous works with $\Delta M_{\rm He-core}\sim 0.006$ \msun\ for all
metallicities (see Cassisi et al. 2007). Neutrino emission due to
plasma processes was calculated as described by Haft et al. (1994). We
used a nuclear network that considers 16 elements and 34 nuclear
reactions of the p-p chains, CNO bi-cycle, helium burning, and carbon
ignition that are identical to those adopted by Althaus et al. (2005)
with the exception of $^{12}$C+$p\rightarrow
^{13}$N+$\gamma\rightarrow ^{13}$C+$e^++\nu_e$ and
$^{13}$C$(p,\gamma)^{14}$N, of special interest in this work, which
are now taken from Angulo et al. (1999).

In this work, mixing and burning of elements are solved simultaneously
in the context of diffusive convective mixing.  This is performed by
solving the set of equations
\begin{equation}
\frac{d\vec{X}}{dt}=\left(\frac{\partial\vec{X}}{\partial t}\right)_{\rm nuc}
+\frac{\partial}{\partial m}\left[\left(4\pi r^2 \rho\right)^2 D
\frac{\partial \vec{X}}{\partial m}\right]
\end{equation}
where $\vec{X}$ is a vector containing the abundances of all considered
 elements (see Althaus et al. 2003 for details of the numerical
 procedure). The efficiency of convection (or any other mixing process) is
 described by adopting the appropriate diffusion coefficient, $D$.  In the
 present work convection was solved (generally, see Sect. 4) with the
 standard mixing length theory (MLT), adopting a value of $\alpha=1.61$ for
 the free parameter of the MLT. This value allowed us to reproduce the
 present luminosity and effective temperature of the sun, $\log T_{\rm
 eff}=3.7614$ and $L_\odot=3.842\times 10^{33}$erg s$^{-1}$, at an age of
 $t=4570$ Myr, when adopting $Z=0.0164$ and $X=0.714$ in agreement with the
 $Z/X$ value of Grevesse \& Sauval (1998).  In the context of the MLT,
 convection is described by
\begin{equation}
D=\frac{\alpha H_P v_{\rm MLT}}{3}=\frac{\alpha^{4/3} H_P}{3}
\left[\frac{cg}{\kappa \rho}(1-\beta) \nabla_{\rm ad}(\nabla_{\rm rad}-\nabla)\right]^{1/3}
\end{equation}

Initial (ZAMS) masses for our simulations (Table 1) were chosen to
obtain ZAHB ages close to those of globular clusters ($\sim 12.5$ Gyr,
Salaris \& Weiss 2002).  However we do not expect initial mass to play
a role in these simulations and the results should therefore be valid
for sequences of ages at the moment of the HeCF similar to that of the
disk ($\sim 8$ Gyr)\footnote{ For example at Z=0.02 the mass of the
He-core at the HeCF is less than 0.5\% smaller for a higher mass
progenitor (1.25 \msun) that undergoes the HeCF $\sim 6$ Gyr after the
ZAMS.}.  Mass loss in {\tt LPCODE} is now included with the prescription of
Schr\"oder and Cuntz (2005). However, as the details of mass loss in
the RGB do not correspond to distinctive features in the models after
the star has departed from the RGB, we enhanced mass loss artificially 
close to the RGB tip to obtain different H-rich envelope masses at the
moment of the HeCF. Mass loss was stopped arbitrarily when the star
reached $T_{\rm eff}\sim12\,600$ K.

The initial helium content of the sequences (which span a wide metallicity
range) was chosen to be 
\begin{equation}
Y=0.23+2.4\times Z,
\end{equation}
 allowing comparison with previous works, that have adopted $Y=0.23$ for
metal-poor globular clusters, while at the same time is consistent with our
adopted solar values and similar to present determinations of the
chemical evolution of the galaxy (Flynn 2004, Casagrande et al. 2007).

\section{Description of standard sequences}
We performed simulations of the late hot-flasher scenario in which the
surface composition is altered by both dilution and burning of the
H-rich envelope. Evolutionary sequences were followed from the ZAMS,
through to the hot-flasher event, and to the post horizontal branch
evolution. Our standard sequences include neither overshooting at any
convective boundary nor any other extra-mixing processes (such as
semiconvection). In these sequences, we also neglected the effect of
the $\mu$-gradients at the H-He transition and convective velocities
were those predicted by the MLT. Since our aim is to provide a
complete and homogeneous set of observable properties of the
predictions of these events, we performed full evolutionary
simulations for the different flavors of the hot-flasher scenario and
four different metallicities (Z=0.001, 0.01, 0.02 and 0.03), and
 for He-core flashes at different points of the post-RGB
evolution, i.e. at different remnant masses. Initial abundances and
masses of these sequences are shown in Table 1.

\begin{table}
\begin{center}
\begin{tabular}{cc}
Mass at ZAMS  &  Initial  Abundances   \\
 (\msun)      & (X/Y/Z)                \\ \hline
0.88          &   0.769/0.230/0.001    \\
0.98          &   0.736/0.254/0.010    \\
1.03          &   0.702/0.278/0.020    \\
1.04          &   0.668/0.302/0.030    \\
\end{tabular}
\label{tab:inicial}
\caption{Initial (ZAMS) values of our standard sequences.}
\end{center}
\end{table}
%As mentioned in the previous section to obtain a helium core flash after the
% star departed from the RGB we artificially enhance mass loss close to the RGB
% tip and, then, shut down when the star becomes hotter than $T_{\rm
% eff}\sim12600$ K.

 Before describing of the surface properties of these sequences, we
 provide a brief description of how H-deficiency is achieved in the
 simulations.

\subsection{Cases of surface H-depletion}
\begin{figure}[ht!]
\begin{center}
\includegraphics[width=180 pt] {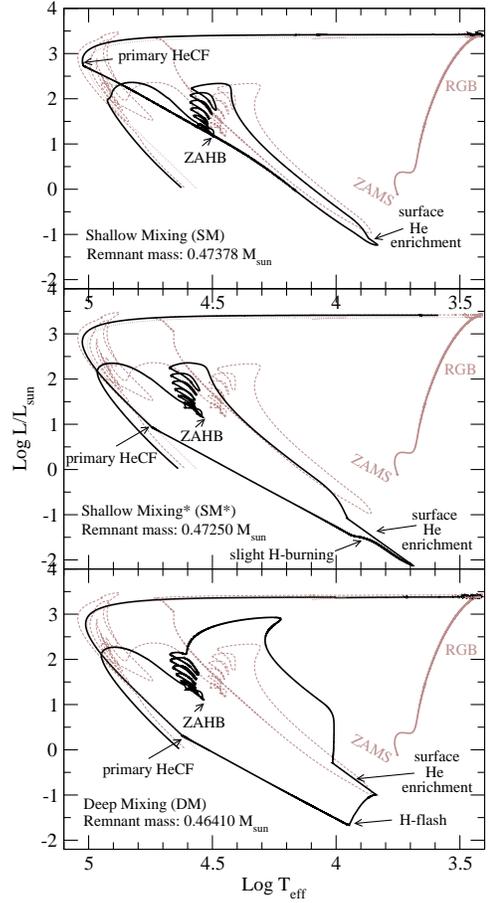}
\caption{HR diagrams for the three kinds of late hot-flashers
described in this work. Sequences correspond to Z=0.02. For
comparison, an early hot-flasher sequence of 0.47426 \msun\ that
finishes with normal H abundances (broken lines), and a HeCF-manque
sequence of 0.4607 \msun\ that ends as a He core white dwarf
(dotted line) are also plotted.
[color figure only available in the electronic version]} 
\label{fig:HRx3}
\end{center}
\end{figure}

Late hot-flashers, i.e. those that experience He-enrichment of their
stellar surfaces, were classified by Lanz et al. (2004) into
``shallow'' and ``deep'' mixing cases. In shallow-mixing (SM) cases,
no H-burning occurs and He-enrichment is attained only by dilution of
the remaining H-rich envelope into deeper regions. In contrast, in
deep-mixing (DM) cases the H-rich envelope is engulfed by He-flash
convective zones and H is burned violently in the hot interior of the
star. This picture is not entirely consistent with our results in
particular because there are intermediate cases in which, although
He-enrichment is due to dilution some inner parts of the H-rich
envelope are burnt during the episode.  In these cases $L_H$, can be
momentaneously as high as $10^7$ $L_\odot$ but only a fraction of the
H-content is finally burnt. Since there is no sharp separation between
these intermediate cases and the ``shallow mixing'' cases of Lanz et
al.  (2004), we retain the ``shallow mixing'' denomination for both
cases. However, we denote the existence of some important H-burning
during the flash by labeling those cases in which H-burning luminosity
reached values momentaneously higher than $10^4\ L_\odot$ as SM$^*$
(see Table 2). On the other hand, the separation between DM and SM
cases is clearly characterized by the development of a violent runaway
burning of protons during DM cases and, especially, by the almost
complete burning of the original H content of the remnant.  In the
following paragraphs, we briefly describe these different cases of
late hot-flashers.

\begin{figure}[ht!]
\begin{center}
\includegraphics[ width=245 pt] {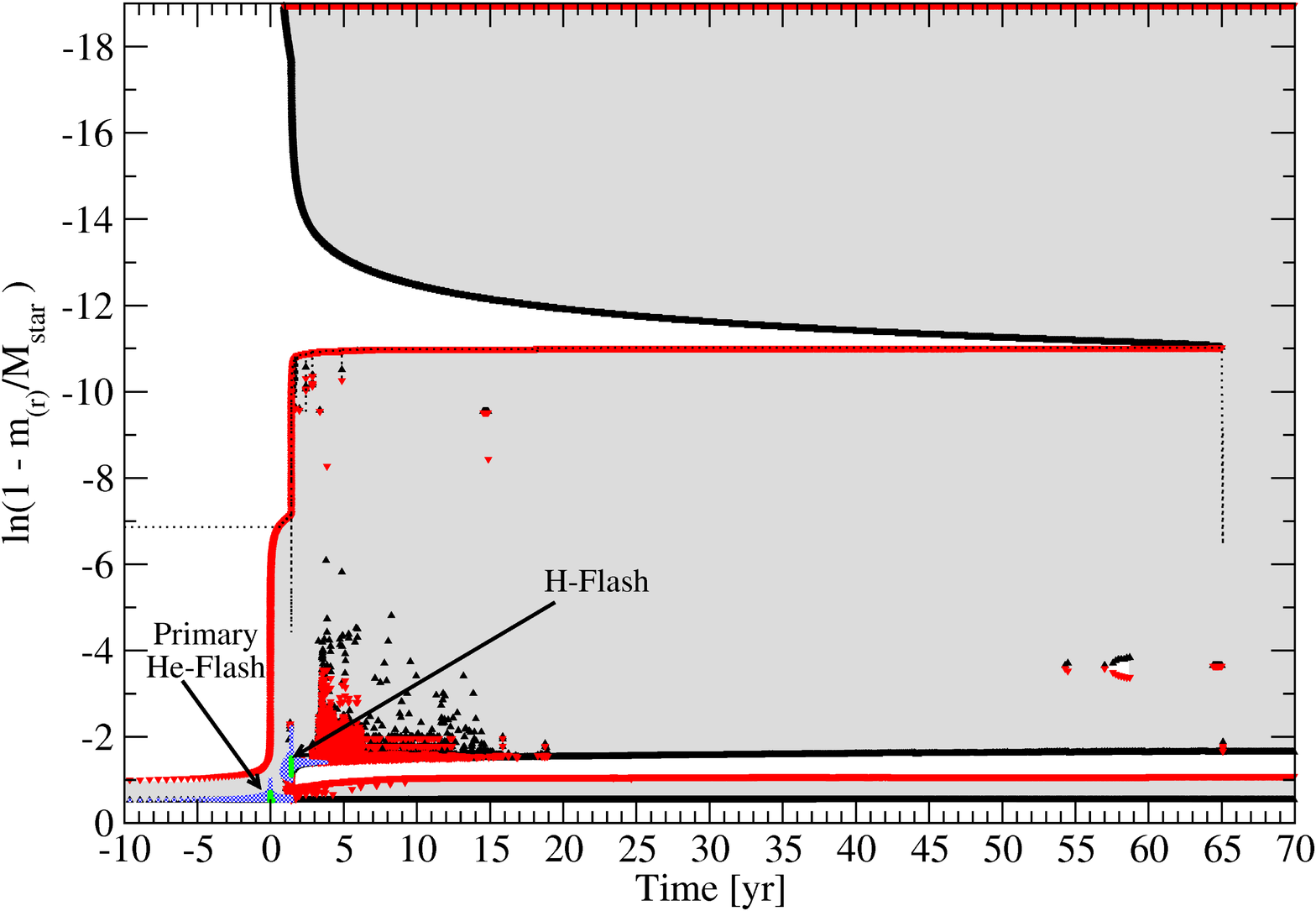}
\caption{Kippenhahn diagram of a deep-mixing case. The case plotted
  corresponds to a 0.47017 \msun\ sequence of initially Z=0.02. Red (black)
  points mark the upper (lower) convective boundaries of each model. Main
  convective zones have been shaded for clarity. Blue (green) zones mark the
  regions in which energy generation per unit mass is above $\epsilon>10^7$
  erg g$^{-1}$ s$^{-1}$ ($\epsilon>10^9$ erg g$^{-1}$ s$^{-1}$). Dotted line
  marks the location of the inner boundary of the H-rich envelope, defined by
  $X \sim 0.001$.
[color figure only available in the electronic version]}
\label{fig:Kipp-DM}
\end{center}
\end{figure}
\begin{figure}[ht!]
\begin{center}
\includegraphics[ width=245 pt] {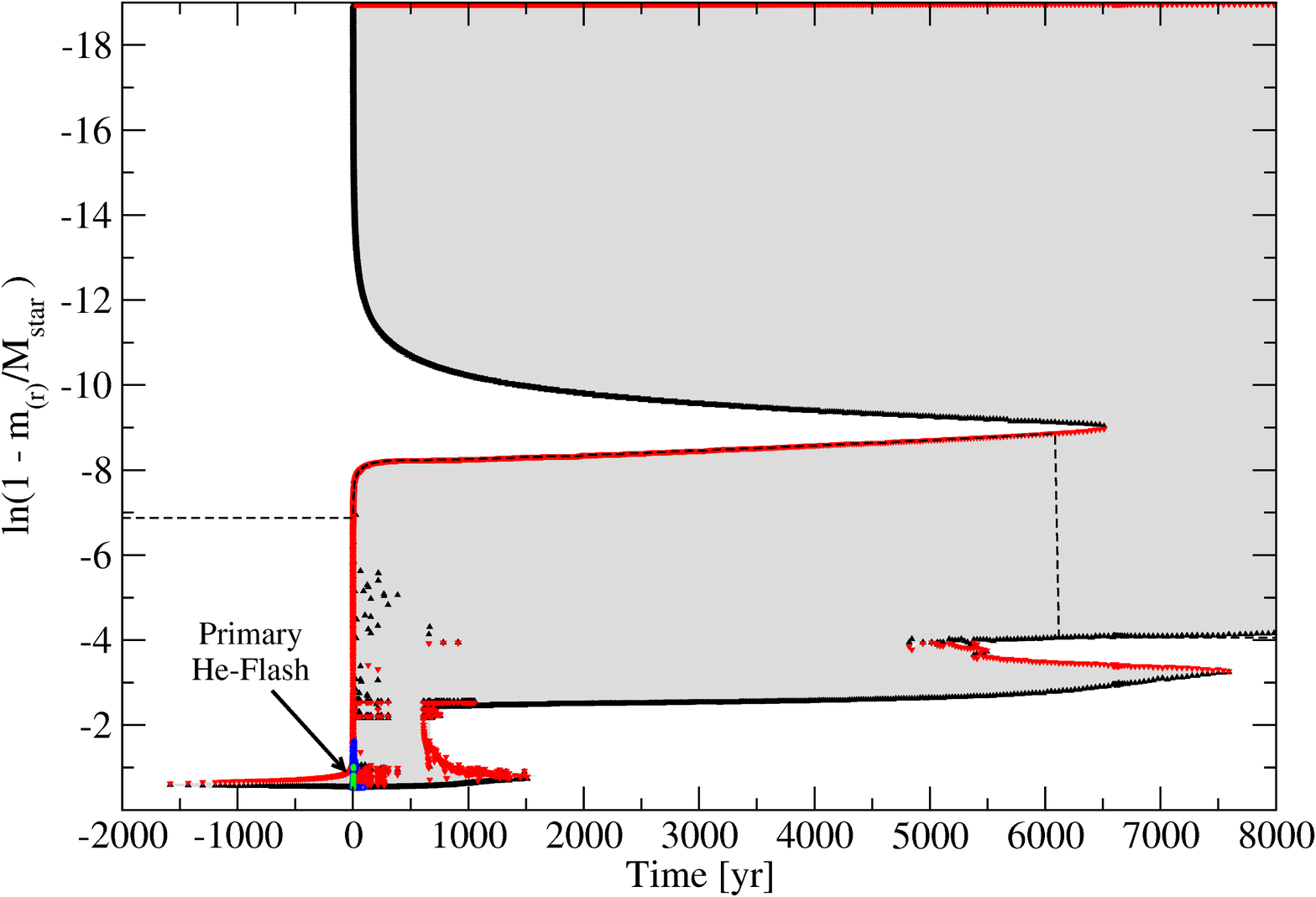}
\caption{Kippenhahn diagram of shallow-mixing with slight CNO burning
  (SM$^*$). References are similar to those of Fig. \ref{fig:Kipp-DM}.
[color figure only available in the electronic version]}
\label{fig:kipp_SM_conCNO}
\end{center}
\end{figure}
\begin{figure}[ht!]
\begin{center}
\includegraphics[width=245 pt] {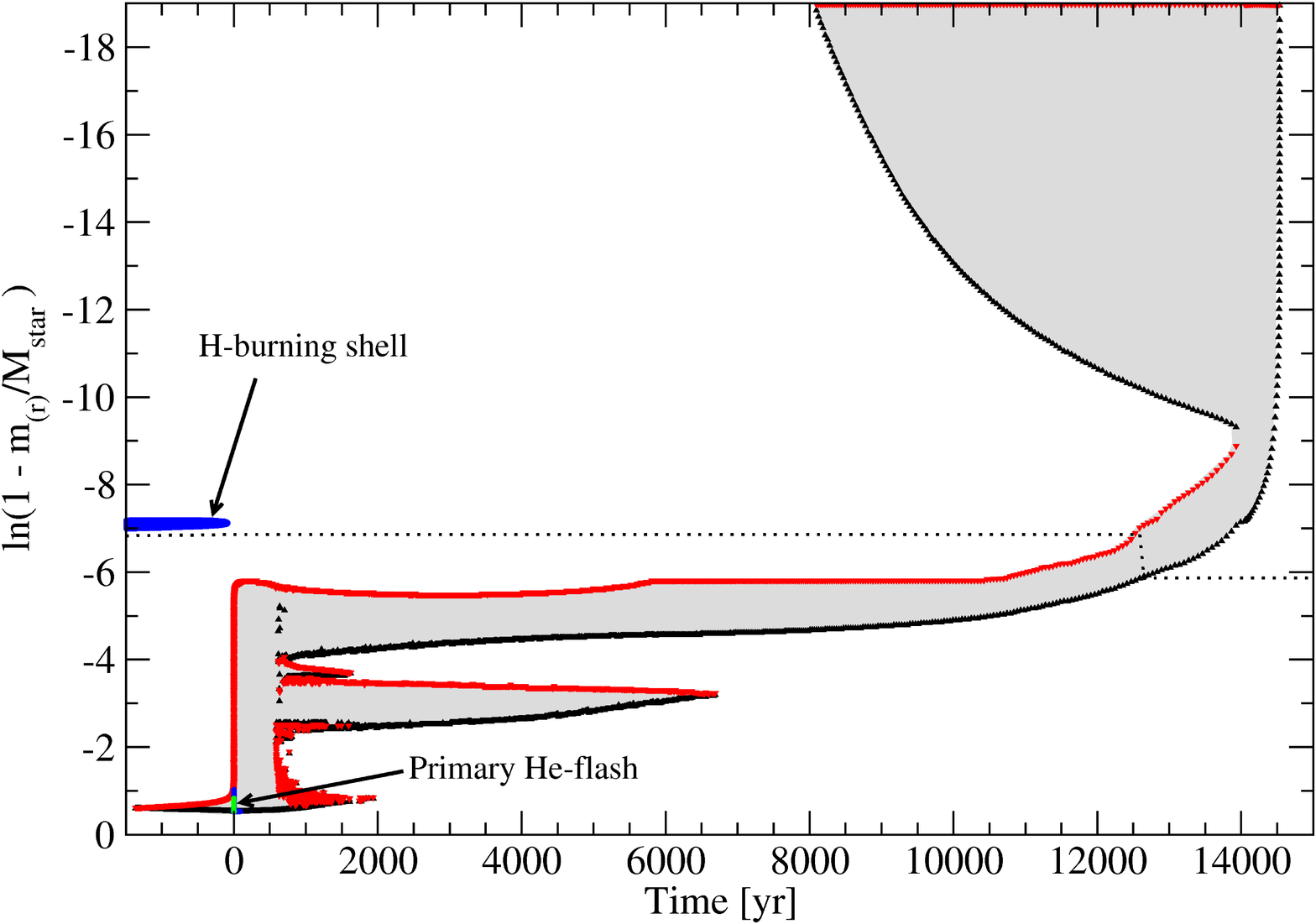}
\caption{Kippenhahn diagram of shallow-mixing without slight CNO
burning. References are similar to those of Fig. \ref{fig:Kipp-DM}.
[color figure only available in the electronic version]}
\label{fig:kipp_SM_sinCNO}
\end{center}
\end{figure}

Since the evolution of convective zones during all cases of late
hot-flashers have some common features, we discuss these first. First,
as a consequence of neutrino cooling, the primary HeCF develops
off-center. A growing convective zone develops in the He-core above
the point of maximum release of energy, with its upper boundary
becoming progressively closer to the H-He transition. When the flash
is fully developed and the structure of the star changes as a
consequence of the sudden energy released by the primary HeCF ---the
star is moving to low luminosity and low temperature regions of the HR
diagram see Fig. \ref{fig:HRx3}--- a very shallow inwardly growing
convective zone develops due to opacity and lower surface temperatures
of the models. It is this convective zone that ultimately exposes the
H-deficient material when it merges with some of the convective zones
developed in the interior of the star. This is true for all kinds of
late hot-flashers (see Figs. \ref{fig:Kipp-DM},
\ref{fig:kipp_SM_conCNO}, and \ref{fig:kipp_SM_sinCNO}).

The evolution of the DM episode is shown in Figs. \ref{fig:HRx3} and
\ref{fig:Kipp-DM}. First, the He-flash starts off-center and similar to
those at the tip of the RGB, developing a convective zone whose upper
boundary becomes closer to the H-He transition as the He-burning
luminosity rises. As reported by Sweigart (1997), the entropy barrier
(due to previous CNO burning) is insufficiently high ---because the
H-burning shell is almost extinguished at the moment of the flash---
and the HeCF driven convective zone is able to penetrate the H-rich
envelope, bringing protons into the C-enriched (a few percent by mass)
He-burning convective zone. As a consequence, H burns producing an
extra energy release. At some point, the rate of H-rich material
brought down and burnt in the interior is so high that the energy
released by H-burning becomes very important and two important events
occur.  First, due to the rise in the entropy of the material at the
location at which protons burn these layers become convectively stable
splitting the previous convective zone into two, one powered by the
He-flash and the other by H-burning. Second, the energy released by
proton burning becomes the main source in driving further expansion
the convective region leading to an unstable situation: a further
increase in the convective zone produces an increase in the power
released by H-burning, which leads to a further outward excursion of
the convective boundary.  This produces a H-flash in which almost all
H-content of the remnant is rapidly burnt, an episode that is also
characterized by a rapid outward excursion of the outer boundary of
the convective zone.  This additional growth in the convective zone
can be clearly seen in Fig. \ref{fig:Kipp-DM} about 1 year after the
primary HeCF. During the episode, the star moves to the low
luminosity, low temperature region of the HR diagram (see
Fig. \ref{fig:HRx3}), where a very shallow outer convective zone
develops. This convective zone becomes progressively deeper and
ultimately, some years after the primary HeCF, merges with the inner
convective zones and the surface of the star becomes H-deficient
(Fig. \ref{fig:Kipp-DM}). After this, the star proceeds through
several secondary helium flashes, like in the case of ``canonical
helium core flashes'' (i.e. at the RGB tip), before finally settling
at the ZAHB as an extremely H-deficient star ($X_H\sim
10^{-4}-10^{-6}$). The hydrogen content decreases from a few
$10^{-4}$\msun\ to usually less than $10^{-6}$\msun\ (see Table 2)
after the primary helium flash.

If the flash occurs when the H-burning shell is somewhat more luminous
(a SM$^*$ case), then the convective zone may come into contact with
the inner part (with low H content) of the H-rich envelope, but the
material that is brought downwards is never sufficient for the
H-flash to develop. This is shown in Fig.
\ref{fig:kipp_SM_conCNO} (see also the middle panel of Fig. \ref{fig:HRx3}).
 Then only a tiny fraction of the envelope is burned ---e.g. the total
H-content of the star changes, after the entire series of HeCFs, from
$2.5\times 10^{-4}$ to $6.9\times 10^{-5}$ \msun, in the case shown in
Fig. \ref{fig:kipp_SM_conCNO}, and in the Z=0.001 SM$^*$ case of Table
2 only about $\sim10^{-4}$ \msun\ is burnt during the primary
HeCF. Afterwards, the outer boundary of the flash driven convective
zone penetrates slowly, diluting the inner regions of the H-rich
envelope until it finally merges with the inwardly growing surface
convective zone, turning the star into a hydrogen deficient star with
low H content in the envelope ($X$ of $10^{-2}$ to $10^{-3}$ by mass
fraction, see Table 3).

Finally, if the flash occurs when the H-burning shell is even more
active (see Fig. \ref{fig:kipp_SM_sinCNO}, and also the top panel of
Fig. \ref{fig:HRx3}), then the convective zone does not come into contact
with the H-rich envelope, due to the entropy, barrier and no H is burned
(this is the case originally referred to as ``shallow mixing'' by Lanz
et al. 2004). After the He-flash, a small tail of the HeCF convective
zone remains for many thousand years moving outwards in mass. The
convective zone reaches the H-rich envelope and dilutes its inner
parts into the H-free interior. Finally, it merges with the
inwardly growing surface convective zone, producing a mild H-deficiency
in the stellar surface.

Results of numerical simulations of all these events are presented in
Table 2, where the timescales and H-content at different stages are given.

\subsection{Sequences in the mass-metallicity plane}

\begin{figure}[]
\begin{center}
\includegraphics[width=240 pt ] {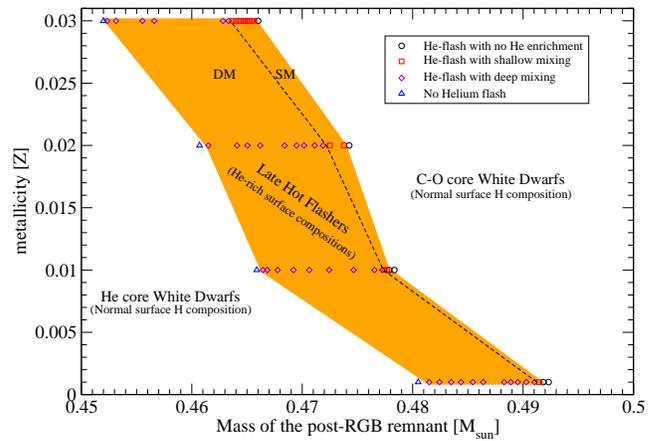}
\caption{Mass and metallicity of the standard sequences calculated for this
  work. The orange zone indicates the sequences that undergo a change
  in their surface abundances as a consequence of mixing and/or
  burning of the H-rich envelope. Dashed line marks the separation
  between DM and SM cases.  [color figure only available in the
  electronic version]}
\label{fig:M-Z}
\end{center}
\end{figure}

 In Fig. \ref{fig:M-Z}, we show the standard sequences calculated for
 this work in the mass-metallicity\footnote{By metallicity we mean the
 original metal (ZAMS) content of the star. During this work we choose
 to label the sequences with their final mass instead of the mass loss
 rate ---as done in previous works--- because it is the final mass of
 the remnant, and not the particular way in which mass is lost, which
 plays an important role in the development of the different kinds of
 hot-flashers.} plane. As previously mentioned the masses of
 hot-flasher remnants are slightly lower than in previous works due to
 the use of the new conductive opacities presented by Cassisi et
 al. (2007).

Our results support the claim by Lanz et al. (2004) that
shallow-mixing is more important in sequences of high metallicity (see
Fig. \ref{fig:M-Z}). While shallow-mixing accounts for only $\sim$4\%
of the remnant mass range of late hot-flashers in lower metallicity
sequences (Z=0.001 and Z=0.01), it corresponds to 14-17\% of the
remnant's mass range in the Z=0.02 and Z=0.03 sequences. The remnant
mass range for hot-flashers is found to be slightly dependent on
metallicity, varying from $\Delta M\sim 0.014$\msun\ for Z=0.03 to
$\Delta M\sim 0.0105$\msun\ for Z=0.001, although this increase is
mainly due to the increase in the SM mass range, while the DM mass
range is almost constant. It is not easy to translate this result into
a mass range for initial masses because this depends strongly on the
mass-loss prescription or mechanism.

\subsection{Surface properties}

\begin{table*}[ht!]
\begin{center}
{\small
\begin{tabular}{c|c||c|c|c|c|c|c|c|c}
Final Mass & Z$_0$ & maximum   &  maximum  & $^1\Delta t$  &  $^2\Delta t$ &
$^3\Delta t$  & $^1$M$_H$ & $^2$M$_H$ & $^3$M$_H$   \\

 [\msun]    &       & Log($L_{\rm He}/L_\odot$) & Log($L_{\rm H}/L_\odot$) &
 [yr]  & [Myr]  & [Myr] & [\msun] & [\msun] & [\msun] \\ \hline\hline
0.49145 (SM$^*$)& 0.001   & 10.23  & 7.44 & 9006 & 1.53 & 66.73
& 4.19$\times 10^{-4}$ & 3.21$\times 10^{-4}$ & 3.21$\times 10^{-4}$ \\
0.49104 (DM)    & 0.001   &  10.22  & 9.24 & 106.78 & 1.55 & 67.33
& 4.17$\times 10^{-4}$ & 2.92$\times 10^{-5}$ & 2.29$\times 10^{-6}$ \\
0.48545 (DM)    & 0.001   & 10.10 & 10.09 & 7.35 & 1.55 & 65.20
& 4.30$\times 10^{-4}$ & 2.37$\times 10^{-7}$ & 4.16$\times 10^{-8}$ \\
0.48150 (DM)    & 0.001   & 10.09 & 10.71 & 4.11 & 1.57 & 69.60
& 4.41$\times 10^{-4}$ & 2.10$\times 10^{-6}$ & 2.46$\times 10^{-7}$ \\\hline
0.47770 (SM)    & 0.01  & 9.83 & 3.58 & 79123 & 1.71 & 71.82
& 2.93$\times 10^{-4}$ & 2.44$\times 10^{-4}$ & 2.44$\times 10^{-4}$ \\
0.47744 (SM$^*$) & 0.01 & 9.83 & 6.91 & 18471 & 1.66 & 71.96
& 2.93$\times 10^{-4}$ & 1.38$\times 10^{-4}$ & 1.31$\times 10^{-4}$ \\
0.47725 (DM)    & 0.01 & 9.82 & 9.12 & 2960 & 1.71 & 72.69
& 2.92$\times 10^{-4}$ & 2.33$\times 10^{-6}$ & 1.85$\times 10^{-7}$  \\
0.46921 (DM)    & 0.01 & 9.66 & 9.91 & 54.7 & 1.71 & 84.69
& 3.06$\times 10^{-4}$ & 6.72$\times 10^{-7}$ & 8.35$\times 10^{-8}$ \\
0.46644 (DM)    & 0.01 & 9.71 & 9.70 & 44.1 & 1.65 & 78.76
& 3.11$\times 10^{-4}$ & 1.96$\times 10^{-7}$ & 3.78$\times 10^{-8}$ \\\hline
0.47378 (SM)     & 0.02 & 9.71 & - & - & 1.80 & 73.89
& 2.62$\times 10^{-4}$ & 2.62$\times 10^{-4}$ & 2.61$\times 10^{-4}$ \\
0.47250 (SM$^*$) & 0.02 & 9.69 & 5.78 & 3053 & 1.80 & 78.99
& 2.54$\times 10^{-4}$ & 9.66$\times 10^{-5}$ & 6.85$\times 10^{-5}$ \\
0.47112 (DM)     & 0.02 & 9.66 & 9.33 & 706 & 1.82 & 76.48
& 2.55$\times 10^{-4}$ & 8.27$\times 10^{-7}$ & 8.95$\times 10^{-8}$\\
0.46410 (DM)     & 0.02 & 9.52 & 10.06 & 135.8 & 1.83 & 88.10
& 2.66$\times 10^{-4}$ & 7.32$\times 10^{-7}$ & 7.91$\times 10^{-8}$\\
0.46150 (DM)     & 0.02 &  9.60 & 10.25 & 132.4 & 1.74 & 81.75
& 2.70$\times 10^{-4}$ & 2.01$\times 10^{-7}$ & 3.48$\times 10^{-8}$\\\hline
0.46521 (SM)    & 0.03 & 9.45 & - & - & 1.87 & 83.33
& 2.67$\times 10^{-4}$ & 2.67$\times 10^{-4}$ & 2.56$\times 10^{-4}$\\
0.46470 (SM)    & 0.03 & 9.45 & 2.67 & 78688 & 1.89 & 79.70
& 2.48$\times 10^{-4}$ & 1.70$\times 10^{-4}$ & 3.44$\times 10^{-5}$\\
0.46367 (SM$^*$) & 0.03 & 9.43 & 5.80 & 17248 & 1.90 & 80.76
& 2.49$\times 10^{-4}$ & 7.23$\times 10^{-5}$ & 1.16$\times 10^{-5}$\\
0.46282 (DM)    & 0.03   &  9.41 & 9.00 & 10926 & 1.93 & 81.47
& 2.49$\times 10^{-4}$ & 4.41$\times 10^{-6}$ & 3.28$\times 10^{-7}$\\
0.45660 (DM)    & 0.03   & 9.27 & 9.80 & 975 & 2.00 & 87.76
& 2.49$\times 10^{-4}$ & 7.05$\times 10^{-7}$ & 5.52$\times 10^{-8}$\\
0.45234 (DM)    & 0.03   & 9.31 & 9.49 & 1055 & 1.94 & 93.93
& 2.58$\times 10^{-4}$ & 5.64$\times 10^{-7}$ & 6.47$\times 10^{-8}$ \\
\end{tabular}
}
\label{tabla}
\caption{Properties of some of the standard sequences calculated for this work.
 Fifth, sixth, and seventh column indicate the interval of time between
 the maximum of Log($L_{\rm He}/L_\odot$) and Log($L_{\rm H}/L_\odot$)
 ($^1\Delta t$), from the primary He flash to the ZAHB ($^2\Delta t$)
 and from the ZAHB to the end of the EHB (TAHB;$^3\Delta t$). The final
 three columns provide the total H content of the sequence just before
 the He flash ($^1$M$_H$), after the primary He core flash ($^2$M$_H$)
 and at the ZAHB ($^3$M$_H$).}
\end{center}
\end{table*}

In Table 3, we present the surface abundances of our simulations for
selected sequences. We first note that, while DM episodes lead to
surface hydrogen abundances preferentially between $10^{-5}$ and
$10^{-6}$ by mass fraction (although they can be as high as a few
$10^{-4}$ in some simulations) the SM/SM$^*$ episodes offer a wide
range of surface H abundances ranging from almost solar to a few
$10^{-3}$ by mass fraction. The SM/SM$^*$ scenarios therefore provide
the intermediate abundances needed by Unglaub (2005)  to
reproduce He-sdB/He-sdO surface abundances in the light of
calculations that include both the effects of diffusion and weak mass
loss rates ($\dot{M}\lesssim 10^{-13}$\msun/yr). This result is quite
interesting because some effects not included
in these simulations (particularly the effect of the $\nabla
\mu-$barrier) may increase the frequency of SM to DM episodes due to
the introduction of an extra barrier against the penetration of the HeCF
convective zone into the H-rich envelope (see Sect. 4).

\begin{table*}[ht!]
\begin{center}
\begin{tabular}{c|c||c|c|c|c|c|c}
Final Mass [\msun]& Z$_0$        & H   & He &  $^{12}$C  &  $^{13}$C
&  N   &  O   \\ \hline\hline
0.49145 (SM$^*$)& 0.001   & 0.2137                & 0.7475 & 0.0365 & $1.57
\times 10^{-6}$ & $1.22 \times 10^{-4}$ & $1.52 \times 10^{-3}$ \\
0.49104 (DM)    & 0.001   & $2.42 \times 10^{-4}$ & 0.9450 & 0.0267 & $7.47
\times 10^{-3}$ & 0.0106 & $4.77 \times 10^{-5}$  \\
0.48545 (DM)    & 0.001   & $4.60 \times 10^{-6}$ & 0.9524 & 0.0264 & $7.73
\times 10^{-3}$ & 0.0134 & $4.53  \times 10^{-5}$ \\
0.48150 (DM)    & 0.001   & $2.41\times 10^{-6}$ & 0.9666 & 0.0107 & $5.79
\times 10^{-3}$  & 0.0185 & $ 3.79\times 10^{-5}$ \\\hline
0.47770 (SM)    & 0.01 & 0.0780 & 0.8682 & 0.0426 & $6.26 \times 10^{-6}$  &
$1.30\times 10^{-3}$ & $6.84\times 10^{-4}$ \\
0.47744 (SM$^*$) & 0.01 & 0.0261 & 0.8996 & 0.0642 & $2.36 \times 10^{-6}$  &
$6.59\times 10^{-4}$ & $7.43\times 10^{-4}$ \\
0.47725 (DM)    & 0.01 & $2.27\times 10^{-5} $ & 0.9464 & 0.0382 & $6.12\times
 10^{-3}$ & $4.35 \times 10^{-3}$ & $2.94\times 10^{-4}$ \\
0.46921 (DM)    & 0.01 & $1.02\times 10^{-5} $ & 0.9560 & 0.0211 & $6.13\times
10^{-3}$ & 0.0128 & $2.92\times 10^{-4}$ \\
0.46644 (DM)    & 0.01 & $4.61 \times 10^{-6}$ & 0.9592 & 0.0193 & $5.52\times
10^{-3}$ & 0.0122 & $3.16 \times 10^{-4}$ \\\hline
0.47378 (SM)     & 0.02 & 0.3822 & 0.5978 & $1.23 \times 10^{-3}$ & $8.96
\times 10^{-5}$ &  $7.57 \times 10^{-3}$ & $4.69 \times 10^{-3}$ \\
0.47250 (SM$^*$) & 0.02 & $9.56 \times 10^{-3}$  & 0.9273 & 0.0425 & $1.93
\times 10^{-5}$ & $2.94 \times 10^{-3}$ & $8.64 \times 10^{-4}$ \\
0.47112 (DM)     & 0.02 & $9.05 \times 10^{-6}$  & 0.9389 & 0.0420 & $ 4.27
  \times 10^{-3}$ & $4.88 \times 10^{-3}$ & $6.96 \times 10^{-4}$ \\
0.46410 (DM)     & 0.02 & $1.05\times 10^{-5}$   & 0.9460 & 0.0273 &
$6.41\times 10^{-3}$ & 0.0120 & $7.680 \times 10^{-4}$ \\
0.46150 (DM)     & 0.02 & $4.66\times 10^{-6}$   & 0.9480 & 0.0263 &
$5.67\times 10^{-3}$ & 0.0116 & $1.01  \times 10^{-3}$  \\\hline
0.46521 (SM)    & 0.03    & 0.2227                & 0.7271 & 0.0197 & $7.79
\times 10^{-5}$ & $8.09\times 10^{-3}$ & $5.18 \times 10^{-3}$ \\
0.46470 (SM)    & 0.03    & $4.71 \times 10^{-3}$ & 0.9239 & 0.0380 &
$4.12\times 10^{-4}$ & $6.10\times 10^{-3}$ & $1.58  \times 10^{-3}$ \\
0.46367 (SM$^*$) & 0.03   &  $1.63 \times 10^{-3}$ & 0.9249 & 0.0390 &
$3.25\times 10^{-3}$ &  $6.48\times 10^{-3}$ & $1.62  \times 10^{-3}$ \\
0.46282 (DM)    & 0.03   &  $7.48 \times 10^{-6}$ & 0.9333 & 0.0420 &
$1.74\times 10^{-3}$ &  $7.04\times 10^{-3}$ & $1.55\times 10^{-3}$  \\
0.45660 (DM)    & 0.03   &  $9.22 \times 10^{-6}$ & 0.9382 & 0.0341 &
$4.09\times 10^{-3}$ & 0.0103 &$1.67\times 10^{-3}$  \\
0.45234 (DM)    & 0.03   &  $1.94 \times 10^{-5}$ & 0.9411 & 0.0273 &
 $4.76\times 10^{-3}$ & 0.0134 &$2.14\times 10^{-3}$
\end{tabular}
\label{tabla_2}
\caption{Surface abundances (by mass fractions) of some of the sequences calculated in
this work.}
\end{center}
\end{table*}

\begin{figure*}[ht!]
\begin{center}
\includegraphics[ width=460 pt] {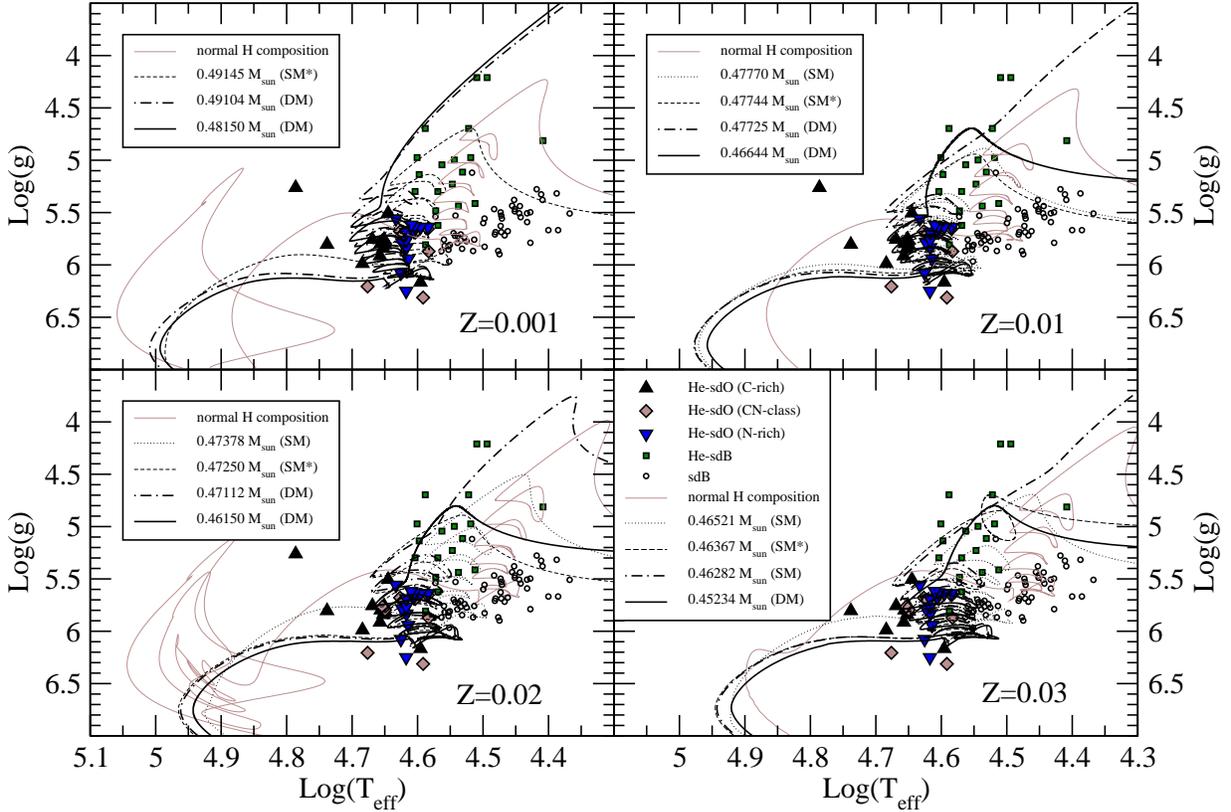}
\caption{  Values of $T_{\rm eff}$ and $g$ of
  our sequences for different metallicities. Values inferred for He-sdO and
  sdB stars are from by Str\"oer et
  al. (2007) and Lisker et al. (2005), respectively. He-sdB data is taken from
  Jeffery (2008) [color figure only available in the electronic version]}
\label{fig:teff-g}
\end{center}
\end{figure*}

For the other element abundances, some trends are evident for the DM
cases (see Table 3). In particular, $^{12}$C decreases by a factor of
2 from earlier (i.e. higher mass remnants) to later cases (i.e.  lower
mass remnants)\footnote{This definition of ``later'' and ``earlier''
does not refer to the time between ZAMS and HeCF but to the location
in the HR-diagram in which the flash takes place.}. This is because,
for later DM episodes, the H-flash and the subsequent splitting of the
convective zone occur more rapidly than in earlier DM cases (see
$^1\Delta t$ values in Table 2) and less $^{12}$C is therefore created
by the time of the H-flash. When the convective zones split, the
increase in $^{12}$C inside the (inner) He-flash driven convective
zone does not affect the final surface abundances.  On the other hand,
$^{13}$C shows no definite trend and is almost constant for a given
metallicity, and decreases slightly when metallicity is increased. In
Table 3, we can see that $^{14}$N clearly increases for later
deep-mixing episodes (for a given metallicity). This can be understood
easily as a consequence of the earlier development, in those cases,
of the H-flash.  Since there is a lower amount of $^{12}$C in the
convective zone when both splitting occurs and H-burning develops, and
the H mass engulfed is almost constant (see Table 2), then there are
more protons per each $^{12}$C nuclei and more $^{14}$N is therefore
formed. Then, at a given metallicity, the ratio of C/N can vary by
more than a factor of two, depending on the moment of the post-RGB
evolution at which the deep mixing occurs.  This is interesting in
view of the two classes of He-sdO stars (N- and C- rich) presented by
Str\"oer et al. (2007). The dependence of the C/N ratio as a function
of initial metallicity indicates that the lower the metallicity, the
higher the N abundance. However, no clear correlation appears to be
present between the location of our sequences with higher N/C ratios
and the N-rich class of Str\"oer et al. (2007) (see
Fig. \ref{fig:teff-g}).

In the case of shallow-mixing events, $^{13}$C is orders of magnitude
lower than for DM because no significant (if any) amount of H burns
during the flash.  At first, it may appear strange that N abundances
in some cases, do not differ significantly (although lower) from those
of DM episodes.  However, during SM events, the material dredged up to
the surface corresponds to material previously processed by the CNO
cycle in the RGB, which transforms almost all primordial C and O into
N. In this case, the N present in the surface is therefore strongly
dependent on the initial metallicity of the stars. This explains why
the N abundance of SM events in Table 3 show a strong dependence on
$Z_0$. The final $^{12}$C abundances in SM depends mainly on the
interplay between the enormous amount of $^{12}$C created by the
primary He core flash and the intensity of the dilution process when
the inner and outer convective zones merge.

 As noted by Str\"oer et al. (2007), although the tracks shown in
Fig. \ref{fig:teff-g} pass close to the location of He-sdO stars
derived by Str\"oer et al. (2005, 2007), their distribution in the
$g$-$T_{\rm eff}$ diagram does not correspond to the one expected from
the simulations.  From the point of view of the models, stars should
cluster towards the location of the EHB, $4.5\lesssim {\rm log}T_{\rm
eff}\lesssim 4.65$ and $5.9\lesssim {\rm log}g\lesssim 6.2$, because
the evolution from the primary HeCF to the settlement in the ZAHB is far
more rapid (occuring in less than $2 \times 10^6$ yr) than the He-core
burning stage ($65-90\times 10^6$ yr, see Table 2).  This clustering
of He-sdO stars is not observed in the inferred $g$-$T_{\rm eff}$
values of real stars (see Fig. \ref{fig:teff-g}).  A possible
explanation for this discrepancy may be that helium-rich atmospheres
become hydrogen-rich, due to diffusion processes, in a timescale
shorter than the duration of the He-core burning stage (EHB). In
Sect. 4.4, we show that such a conversion is, in fact,
possible. According to this scenario, late hot-flashers have a He-rich
surface composition only during the evolution towards the EHB and then
become stars with hydrogen dominated atmospheres. This implies that
He-sdO stars evolve into hydrogen-rich sdB/sdO stars, so that no
clustering of He-sdO stars around the EHB is expected. This scenario
is discussed in Sects. 4.4 and 5.

When comparing theoretical predictions with the results of Str\"oer et
al. (2007), we recall that surface properties in that work were
obtained by adopting H/He atmospheres, which was found by Lanz et
al. (2004) (who adopted H/He/C atmospheres in their analyzes) to be a
possible cause of systematic effects.  The effective temperature
inferred in real stars may also be systematically overestimated due to
line-blanketing effects (see Str\"oer et al. 2007).

 Finally, we note that the surface characteristics of our solar
metallicity DM sequences agree closely with the values inferred by
Lanz et al. (2004) for PG1544+488. This result clearly confirms the
statement of Lanz et al. (2004) that this object could be the result
of a DM event. Our results, also, confirm that the surface abundances
derived by these authors for JL87 agree with those predicted by the SM
kind of hot-flasher. In contrast, no sequence presented abundances
close to those of LB1766, confirming that this object may have had a
very different history.

\subsection{Comparison with previous works}
We compare our results with the very few consistent simulations available
in the literature. In the DM, we compare the 0.49125 sequence of Cassisi et
al. (2003), for which the He-core flash ocurrs at $\log L/L_\odot\sim 0.79$ (a
relatively ``early'' deep-mixing episode in the terminology of Sect.
3.2) with the most similar models in our grid. This is shown in Table \ref{comparacion}
\begin{table}[ht!]
\begin{center}
\begin{tabular}{c||c|c|c}
             & this work & this work & Cassisi et al.  \\\hline Mass [\msun] &
0.49104 & 0.48545 & 0.49125 \\ $Z_0$ & 0.001 & 0.001 & 0.0015 \\ $^1\Delta t$
& 107 d & 7.35 d & 150 d \\ max. $L_{\rm He} /L_\odot$ &1.66 $\times 10^{10}$
& 1.26 $\times 10^{10}$ & 3.7 $\times 10^{10}$ \\ max. $L_{\rm H} /L_\odot$
&1.74 $\times 10^{9}$ & 1.23 $\times 10^{10}$ & 8.2 $\times 10^{9}$ \\ M$_H$
burnt [\msun]& 3.88 $\times 10^{-4}$& 4.30 $\times 10^{-4}$& 3.7 $\times
10^{-4}$ \\ surf. H & 2.4 $\times 10^{-4}$ & 4.6 $\times 10^{-6}$ & 4 $\times
10^{-4}$ \\ surf. He & 0.945 & 0.952 & 0.96 \\ surf. C & 0.0342 & 0.0341 &
0.029 \\ surf. N & 0.0106 & 0.0134 & 0.007 \\ surf. O & 4.8 $\times 10^{-5}$ &
4.5 $\times 10^{-5}$ & 3.5 $\times 10^{-5}$
\end{tabular}
\caption{Comparison of characteristics of DM sequences of this work with the
  DM sequence of Cassisi et al. (2003).}
\label{comparacion}
\end{center}
\end{table}
where the results for the 0.48545 \msun\ and 0.49104 \msun\ (with
Z=0.001) sequences of this work are compared with the DM sequence of Cassisi et
al. (2003). We note in particular the good agreement between the earlier
deep-mixing sequence (0.49104 \msun) and Cassisi et al. (2003) values,
regardless of the different metallicity and mass in both sequences.

For the SM case, the only available sequence is one of solar
 metallicity discussed by Lanz et al. (2004), which we can compare
 with our 0.47378 \msun, Z=0.02 SM sequence. In their sequence the
 shallow-mixing event (the merging of the inner and outer convective
 zones) occurs about $24\,000$ yr after the primary He-core flash
 compared with the $14\,000$ yr in our sequence (see Fig. 3). The
 surface abundances in their sequence are He/C/N=0.5/0.008/0.005
 qualitatively similar to ours (He/C/N=0.6/0.001/0.008). We can
 therefore conclude that there is qualitative agreement between Lanz
 et al.  (2004) and our results.

\section{Alternative sequences}
In this section we discuss the effect of some deviations from the
standard assumptions made in the previous section.  In particular, we
address the effect of possible extra-mixing at convective
boundaries, different mixing velocities, and chemical
gradients. Finally, we address the effect of diffusion in the
outer layers of the standard sequences.

\subsection{Effect of the location of proton burning (altering $D$)}
The calculation of 1D-stellar evolution models through the H-flash in
previous simulations (Sect. 3) were performed within the picture of
diffusive convective mixing. This approach was used in the past for
similar situations like late helium-shell flashes (i.e.  born again
episodes) experienced by post-AGB stars at the beginning of their
white dwarf stage (e.g. Iben \& Mac Donald 1995). This picture, which
implies that material is homogeneously mixed in each mass shell, is
very probably far from reality (see Woodward et al. 2008 for a
discussion of this issue) in which convection appears to be an
advection process. We now analyze how deep-mixing episodes can be
altered if the location of maximum energy generation is displaced from
the predictions of MLT plus diffusive mixing. This is done by altering
the value of the convective diffusion coefficient $D$ in the mixing
equations. We note, however, that because we still assume horizontal
homogeneity, these results can only be considered poor approximations
of the problem\footnote{Horizontal inhomogeneities in $^{12}$C,
$^{13}$C and H would certainly affect the final $^{14}$N yield.  This
is exactly the situation expected during the hydrogen flash (see
Woodward et al. 2008).}.

\begin{table*}[ht!]
\begin{center}
{\small
\begin{tabular}{c||c|c|c|c|c|c||c|c}
Diff. Coef.    & H   & He &  $^{12}$C  &  $^{13}$C &  N   &  O  & $^2$M$_H$ & $^3$M$_H$  \\ \hline\hline
$^\dagger D_{MLT}\times 10$   & $1.03 \times 10^{-3}$ & 0.9022 & 0.0942 & $2.09 \times 10^{-6}$
& 1.08 $\times 10^{-6}$ & $ 4.09  \times 10^{-4}$
& 1.42$\times 10^{-5}$ & 8.12$\times 10^{-6}$ \\
$D_{MLT}\times 3$   & $1.71 \times 10^{-6}$ & 0.9514 & 0.0283 & $7.99 \times 10^{-3}$
& 0.0118 & $5.13  \times 10^{-5}$
& 8.18$\times 10^{-8}$ & 1.54$\times 10^{-8}$ \\
$D_{MLT}$   & $4.60 \times 10^{-6}$ & 0.9524 & 0.0264 & $7.73 \times 10^{-3}$
& 0.0134 & $4.53  \times 10^{-5}$
& 2.37$\times 10^{-7}$ & 4.16$\times 10^{-8}$ \\
$D_{MLT}/10$   & $1.66 \times 10^{-5}$ & 0.9532 & 0.0291 & $7.50 \times 10^{-3}$
& $9.71 \times 10^{-3}$  & $3.62  \times 10^{-5}$
& 7.97$\times 10^{-7}$ & 1.49$\times 10^{-7}$ \\
$D_{MLT}/100$   & $4.22 \times 10^{-5}$ & 0.9506 & 0.0328 & $7.66 \times 10^{-3}$
& $8.49 \times 10^{-3}$  & $3.18  \times 10^{-5}$
& 1.95$\times 10^{-6}$ & 3.72$\times 10^{-7}$ \\
$D_{MLT}/1000$   & $4.57 \times 10^{-4}$ & 0.9515 & 0.0327 & $6.89 \times 10^{-3}$
& $7.99 \times 10^{-3}$  & $3.06  \times 10^{-5}$
& 2.49$\times 10^{-5}$ & 3.95$\times 10^{-6}$ \\
$D_{MLT}/10000$   & 0.0117 & 0.9455 & 0.0376 & $2.53 \times 10^{-3}$
& $1.87 \times 10^{-3}$  & $3.35  \times 10^{-5}$
& 2.27$\times 10^{-4}$ & 7.42$\times 10^{-5}$ \\
\end{tabular}
}
\label{tab:Abu-D}
\caption{Surface abundances (by mass fractions) of the sequences
calculated in this work with different mixing efficiencies.  All
sequences correspond to models with a final mass equal to
0.48545\msun\ and Z=0.001. $^\dagger$This sequence did not undergo the
H-flash, see text.}
\end{center}
\end{table*}

Surface abundances in simulations with reduced mixing efficiencies are
shown in Table 5. Our results are qualitatively consistent with the
preliminary results quoted by Cassisi et al.  (2003). They indicate,
however, that a reduction in $D$ of only two orders of magnitude is
sufficient to produce an increase in the residual amount of H after a
DM episode of about an order of magnitude. In view of the results of
Unglaub (2005), it is interesting to note that a reduction in $D$ of a
factor of between 1000 and 10000, produces H-abundances after the DM
episode that would finally convert He-sdO stars into He-deficient sdB
stars by gravitational settling on the EHB, even in the presence of
weak winds ($\dot{M}\sim 10^{-13}$\msun/yr).  If major proton burning
is shifted deeper into the star ($D\times 10$) then the extra energy
released by proton burning is locally less relevant ---in comparison
with the local flux due to the primary He-flash--- and the violent
proton burning is never sufficient to drive convection further out. As
a consequence, the ``runaway'' burning of protons does not develop
and, as can be seen from Table 5, far higher H abundances are left
than in the standard case. The H-abundance after the primary He-flash
in this case, may also lead to the conversion of He-sdO stars into
He-deficient sdB stars, according to Unglaub (2005). We note also that
a feature of the absence of the runaway proton burning (H-flash) is
that the convective zone developed by the primary HeCF never splits
and the surface $^{12}$C abundance continues to increase until the
convective zone merges with the outer convection. In this case, the
final $^{12}$C is much larger than the cases in which the H-flash
develops.

\subsection{Effect of extra-mixing at convective boundaries}

To analyze the effects of possible extra-mixing episodes, we included
extra-mixing zones at each convective boundary following the
prescription of Herwig et al. (1997). We therefore allowed
extra-mixing episodes by considering that the mixing velocities decay
exponentially beyond each convective boundary. The diffusion
coefficient beyond the formal convective boundary was then given by
\begin{equation}
D_{\rm OV}(z)=D_0 \exp\left(\frac{-2z}{f\,H_P}\right)
\end{equation}
where $D_0$ is the diffusion coefficient of unstable regions close to
the convective boundary and $z$ is the geometric distance from the
edge of the convective boundary. We performed simulations
with three different values of the free parameter $f$ ($f$=0.005,
0.01 and 0.1) around the value inferred from the width of the
upper main sequence ($f\sim 0.016$) and oxygen abundances of
PG1159 stars (Herwig 2000).

Results of these numerical experiments are shown in Table 6.  In the
case of $f=0.005$, the final surface abundances are close to those of
the standard simulations in which no extra-mixing was considered. For
larger extra-mixing efficiencies H-burning becomes more efficient and for
values beyond $f=0.01$ almost no H remains in the star after the
H-flash. For values of $f\lesssim 0.01$, the evolutionary timescales
($^1\Delta t$, $^2\Delta t$ ,$^3\Delta t$;  see Table 2 for the
definition) are unaffected and are similar to the standard case
(see Table 2). In contrast, for the $f=0.1$ case, the time interval
between flashes is significantly reduced (46 d) and the EHB evolution
is somewhat enlarged by the appearance of breathing pulse
instabilities at the end of the EHB (212 Myr). The lower N abundance in
the $f=0.1$ case is because the H is mixed with a larger mass of
$^{12}$C reducing the amount of protons per carbon nuclei.

\begin{table*}[ht!]
\begin{center}
{\small
\begin{tabular}{c||c|c|c|c|c|c||c|c}
$f$-value    & H   & He &  $^{12}$C  &  $^{13}$C &  N   &  O  & $^2$M$_H$ & $^3$M$_H$  \\ \hline\hline
0.005   & $2.48 \times 10^{-4}$ & 0.9540 & 0.0220 & $4.95 \times 10^{-3}$
& 0.0114 & $5.88  \times 10^{-4}$
& 3.14 $\times 10^{-6}$ & 1.93$\times 10^{-6}$ \\
0.01   & $1.18 \times 10^{-8}$ & 0.9548 & 0.0216 & $4.88 \times 10^{-3}$
& 0.0114 & $5.60  \times 10^{-4}$
& 5.05$\times 10^{-10}$ & 9.128$\times 10^{-11}$ \\
0.1   & $1.52 \times 10^{-10}$ & 0.9597 & 0.0227 & $8.41 \times 10^{-5}$
& $7.31 \times 10^{-3}$  & $2.61  \times 10^{-3}$
& 6.85$\times 10^{-12}$ & 1.21$\times 10^{-12}$ \\
\end{tabular}
} \label{tab:Abu-f} \caption{Surface abundances of sequences in which
extra-mixing at convective boundaries was considered with different
efficiencies ($f$).  The sequences correspond to models with a
final mass equal to 0.4641\msun\ and Z=0.02.}
\end{center}
\end{table*}

\subsection{Effect of chemical gradients ($\nabla\mu$)}
Miller Bertolami et al. (2006) recently showed that chemical gradients
($\nabla \mu$) can affect the development of the rapid ingestion of
protons into convective zones driven by helium shell flashes. It
therefore appears important to analyze the possible effect of $\nabla
\mu$ in deep-mixing events. To analyze the effect of
chemical gradients during deep-mixing events, we  performed
some simulations of deep-mixing for a model of 0.4641
\msun\ and Z=0.02 with the theory of convection developed by Grossman
et al. (1993), in its local version presented by Grossman \& Taam
(1996)\footnote{ For simulations adopting the convection theory of
Grossman et al. (1993), the value of the free parameter $\alpha$ is
reduced by $^4\sqrt{2}$ from that adopted in the standard MLT due to a
difference in the definitions of the relevant coefficients.}. This 
theory incorporates the effect of $\nabla \mu$ both in the
determination of the different mixing regimes and  the corresponding
mixing velocities.

This set of simulations\footnote{We performed several simulations
  under different numerical assumptions.} showed that the inclusion of
  $\nabla \mu$ prevents the complete penetration of the He-core flash
  convective zone into the regions of high H-content. In particular,
  at the H-He transition, the convective zone splits into different
  convective zones and, consequently, the chemical profile at the H-He
  transition acquires a staircase-like aspect. Eventually some of
  these convective zones merge and some H can be burnt. However, the
  main consequence is that in these simulations almost all the H-rich
  envelope is not engulfed and burned. In all of these simulations the
  final abundances remain unchanged after the flash, because even the
  dilution event is prevented due to the $\mu$-barrier.

However, the theory adopted here is a local theory and does not
include any kind of extra-mixing beyond the formal convective
boundaries, such as those produced by overshooting (Freytag et al.
1996) or the excitation of gravity waves (Herwig et al. 2006,
2007). Our simulations indicate that the inclusion of an overshooting
efficiency as small as $f=0.001$ in our models allows the
penetration of the $\mu$-barrier and the development of the
H-flash, the surface abundances that arise in these cases are
similar to those in which $\mu$ effects are not taken into
account. Lower $f$-values are difficult to test in our simulations
because they require very high spatial resolution. Although
hydrodynamical simulations of convection in homogeneous mediums
appears to support the existence of some extra-mixing at convective
boundaries, it should be noted that overshooting is expected to be
reduced by the $\mu$-barrier itself (Canuto 1998).

\subsection{Effect of diffusion in the outer layers}

\begin{figure}[ht!]
\begin{center}
\includegraphics[width=230 pt] {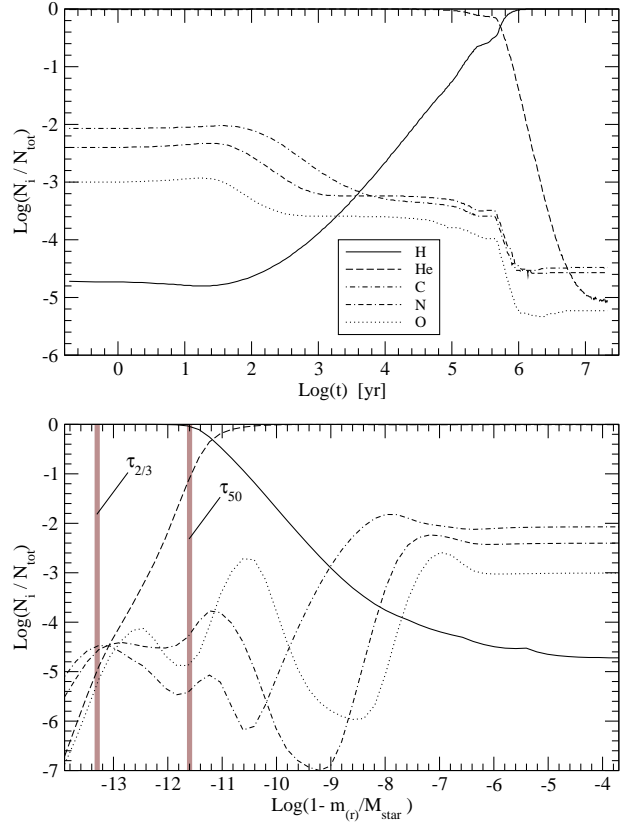}
\caption{Top: Time evolution of the chemical abundances 
at $\overline \tau=2/3$ as a consequence of gravitational settling and
radiative levitation. Initial abundances correspond to one of our DM
sequences (Table 2; M$=0.48545$\msun\ and Z=0.001) with surface
parameters $T_{\rm eff} = 40\,000$ K and log $g =6.0$. Bottom: Chemical
profile of the outer layers after $10^{7}$ yr when the surface was
already H-dominated. Note the very steep chemical profile close to the
photosphere ($\tau_{2/3}$) which can strongly affect the surface
values if some mixing process is effective close to the
photosphere.  In particular note the supersolar He-abundances below
$\overline \tau = 50$ ($\tau_{50}$), not far from the location where the
bottom of the convective zone due to the  He-ionization should be located.}
\label{fig:chem_t}
\end{center}
\end{figure}

We now address the possible effects of the interplay between
mass-loss, gravitational settling, chemical diffusion, and radiative
levitation in our standard sequences (Tabs. 2 and 3). It is well known
that the abundances of sdB stars depend strongly on the combined
effects of diffusion and mass-loss (Unglaub \& Bues 2001), and the
same should be true for He-rich subdwarfs (Unglaub 2005). If mass loss
is higher than $10^{-12} M_{\odot} / \rm yr$, diffusion would not be
able to alter the post-HeCF abundances significantly and the star
would remain He-rich during the entire EHB evolution. For the low
hydrogen abundances of our DM sequences, a mass-loss rate of
approximately $10^{-14} M_{\odot} / \rm yr$ would be sufficient to
prevent transformation of a He-rich into a H-rich star due to the
upward diffusion of hydrogen.

From the theoretical predictions of Vink \& Cassisi (2002) and Unglaub
(2008), it follows that for stars with $M_{*} \approx 0.5 M_{\odot}$,
$T_{\rm eff} = 40\,000 \rm K$, $\log g \la 5.5$, and solar composition,
mass-loss rates of $\dot M \ga 10^{-11} M_{\odot} / \rm yr$ are
expected. The effect of diffusion should therefore be negligible.  For
 more compact stars with $\log g > 5.5$, from the theory of
radiatively-driven winds, which neglects multi-component effects, weak
winds with $\dot M < 10^{-11} M_{\rm \odot} / \rm yr$ are
predicted. However, the results of Unglaub (2008) showed that those
weak winds cannot be ``chemically homogeneous'' as assumed in previous
investigations. Hydrogen and helium, at least,  can hardly be expelled
from the star. Therefore, with the method described in Unglaub \& Bues
(1998), we performed simulations, in which the absence of any
mass-loss was assumed.

\begin{figure*}[ht!]
\begin{center}
\includegraphics[width=300 pt] {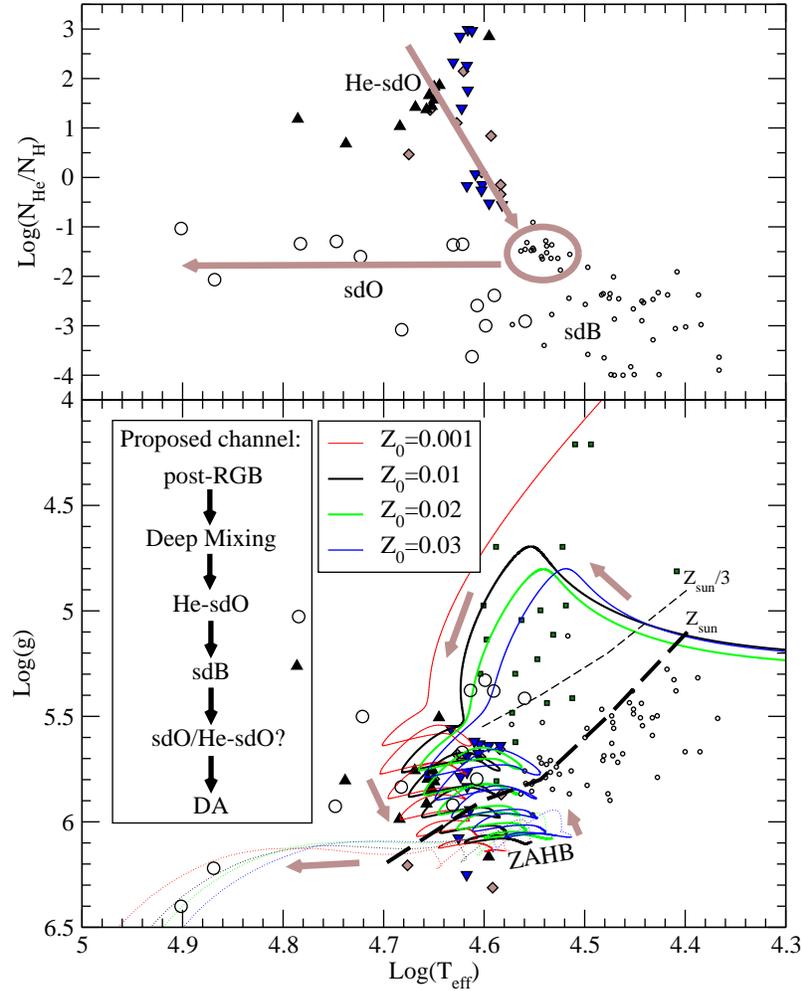}
\caption{Sketch of the proposed channel for 
the evolution after a DM event as a consequence of chemical
diffusion in the absence of winds (thick arrows mark the sense of
evolution). A similar situation would be expected for post-SM/SM$^*$
sequences with weak winds ($\dot{M}\sim 10^{-13}$\msun/yr). Upper
panel: proposed evolution in the abundances-$T_{\rm eff}$ plane,
abundances taken from Str\"oer et al. (2007) and Lisker et al
(2005). Lower Panel: Evolution in the $g-T_{\rm eff}$
plane. Continuous lines correspond to the evolution as a He-rich
subdwarf, while dotted lines mark the evolution after the ZAHB when
diffusion would have turned the atmosphere into a H-dominated
one. Black dashed lines show the wind limit for solar-like abundances
and for two different metallicities (thick line: $Z_\odot$, thin line:
$Z_\odot/3$). [color figure only available in the electronic version]}
\label{fig:channel}
\end{center}
\end{figure*}

In the upper panel of Fig. \ref{fig:chem_t}, we show the evolution of
the surface abundances (at a Rosseland mean optical depth $\overline
\tau = 2/3)$ predicted by one of our standard DM
sequences\footnote{With the exception of the O abundance that was
defined to be $\log(N_{\rm O}/N_{\rm tot})=-3$, instead of
$\log(N_{\rm O}/N_{\rm tot})\sim-5$ as follows from our simulations,
for numerical reasons.  For an initial value of $\log(N_{\rm O}/N_{\rm
tot})\sim-5$, oxygen moves outward in the outer regions and its
abundance tends to its equilibrium value, which is higher than
$\log(N_{\rm O}/N_{\rm tot})\sim-5$. In the underlying regions,
however, where oxygen has helium-like configuration, the outward
radiative force is negligibly small and oxygen sinks. For an
initial abundance of $\log(N_{\rm O}/N_{\rm tot})\sim-5$, this
situation leads to extremely steep concentration gradients in these
regions where the flow changes its sign and, as a consequence, to
numerical problems.}
%----------------------------------------------
 (that with $0.48545 M_{\odot}$ and $Z = 0.001$) as a
consequence of gravitational settling and radiative levitation (at
$T_{\rm eff} = 40\,000
\rm K$ and $\log g = 6.0$). In the absence of mass loss, 
the star changes from a He-rich subdwarf into a He-deficient one in only
$\sim 10^{6} \, \rm yr$. Since this timescale is similar to the time it
takes the star to reach the ZAHB, surface abundances are expected to
become H-dominated as the star settles onto the ZAHB. This leads to the
evolutionary channel sketched in Fig.  \ref{fig:channel}., in which
He-sdO stars become hot sdB stars due to element diffusion.
After about $10^{7} \rm yr$, the surface abundances are close to the
values expected from the equilibrium condition between gravitational
settling and radiative levitation obtained by Unglaub \& Bues
(1998). This timescale is significantly shorter than the timescales
for stellar evolution close to the EHB ($65$ to $90 \rm Myr$). For  higher
$\rm H / \rm He$ ratios, as predicted by our shallow-mixing sequences,
the stellar atmosphere becomes  hydrogen-rich on even shorter
timescales. For the abundances from the SM$^*$ sequence with
$0.47250 M_{\odot}$ and $Z=0.02$, the equilibrium abundances should
already be reached by $10^{5} \rm yr$. In the absence of
mass loss or another disturbing processes, the surface composition of
the majority of sdB stars should therefore be given by the equilibrium
condition between gravitational settling and radiative levitation,
independently from the evolutionary scenario from which they originate.

It is well known that the equilibrium abundances of helium are lower
than observed (Michaud et al. 1989). The number ratios $\rm He /
\rm H$ of the hottest sdB stars are approximately $0.01$ (e.g. Lisker et
al. 2005). There is therefore a discrepancy of about three orders of
magnitude between the predictions and the observations. In
Fig. \ref{fig:chem_t}, we show the predicted abundances as a function
of depth after $10^{7}$ yr. We note the very steep chemical gradient in
the outer regions. Within the stellar atmosphere, the helium abundance
increases from $\rm He / \rm H \approx 10^{-5}$ at $\overline
\tau = 2/3$ to supersolar values at $\overline \tau \ga 50$ (see Fig. \ref{fig:chem_t}). We therefore suggest
that the presence of even weak mixing processes (e.g. due to stellar
rotation), which tend to smooth concentration gradients, should
produce higher helium abundances close to the stellar surface. Thus
the large discrepancy for the helium abundance does not necessarily
exclude the proposed scenario. In future investigations, it should be
shown how mixing processes other than mass loss would affect the
upward diffusion of hydrogen, and the gravitational settling of helium
in a H-dominated stellar atmosphere. In particular, the possible
effect of a thin, outer, convection zone due to changes in ionization
of helium, predicted by Groth et al. (1985) at least for supersolar
helium abundances, could affect the composition of the stellar
atmosphere. For example, in the DM sequence analyzed in this section,
stellar evolution models (without diffusion) show that the bottom of
this convective region should be located around $\log
(1-m/M_\star)\sim -12$ ---with the exact location depending on the
instantaneous $\log T_{\rm eff}$ and $\log g$ values--- where He
abundance is significantly higher (see Fig. \ref{fig:chem_t}).
  
In the $\log T_{\rm eff}$ - $\log g$ diagram of Fig.
\ref{fig:channel}, the line is shown above which, according to Unglaub
(2008), chemical homogeneous winds may exist for stars of solar
metallicity. For stars above this line, the mass-loss rates should be
sufficiently high to prevent diffusion. The chemical composition
should then depend essentially on their evolutionary history. For
stars below this ``wind limit", no chemically homogeneous winds can
exist. The surface composition of these stars should essentially
depend on atmospheric processes such as gravitational settling and
radiative levitation. Pure metallic winds may possibly still exist,
which could produce additional changes in the metal abundances. Due to
the complexity of the situation, no reliable quantitative predictions
appear to be presently possible. Our results in Fig. \ref{fig:chem_t}
indicate, however, that in the absence of mass-loss a helium-rich
atmosphere may become a hydrogen-rich one, even for the low hydrogen
abundances predicted by the DM sequences.

From Fig.  \ref{fig:channel}, it can be seen that our DM sequences
cross the wind limit before the star settles onto the ZAHB. We then
expect the onset of diffusion. A similar situation occurs in hot white
dwarfs on the upper cooling sequence (Unglaub \& Bues 2000). As soon
as winds cease, the abundances of heavy elements decrease and
helium-rich white dwarfs became hydrogen-rich. We emphasize that the
location of the wind limit in the $\log T_{\rm eff}$ - $\log g$
diagram depends on the chemical composition and the line shown in
Fig.  \ref{fig:channel} is for solar abundances. It would be
necessary to investigate the dependence of the mass-loss rates on the
abundances of individual elements. From Fig.
\ref{fig:channel}, it can be seen that there is at least qualitative
agreement between our predictions and observational results.  The
helium-poor sdB stars are preferably in the region below the wind
limit, where gravitational settling could be effective. The He-rich
subdwarfs are preferably above the wind limit, where we expect no
diffusion. The composition of these stars should show clear signatures
of their evolutionary history.

\section{Discussion}

In Sect. 3.3, we showed that standard sequences lead to final H
abundances (see Table 3) immediately after a DM event that are much
lower than inferred for He-sdO stars by Str\"oer et al. (2007). Only a
minority of standard sequences (those which undergo SM/SM$^*$ events)
display abundances similar to He-sdO stars immediately after the
flash.  Also, as we have already mentioned, the predicted location in the
$g-T_{\rm eff}$ would differ from that observed. In
particular, since the EHB evolution is more than 40 times longer than the
evolution from the primary HeCF to the ZAHB, most of the stars
should be clustered around the EHB, which is not
observed. However, there are some viable scenarios in which our
results are consistent with observations. These scenarios
depend on the existence or not of weak homogeneous winds in He-rich
subdwarf stars as we now discuss.

As mentioned in Sect. 4.4, in a very recent study Unglaub (2008)
showed that weak homogeneous winds are impossible in hot subdwarfs at
high gravities. In this case, as shown in Fig. \ref{fig:chem_t}, the
low amount of H remaining in our standard sequences after a DM event
is sufficient to convert a He-rich to a He-deficient surface
composition in about $10^6$ yr. Within this picture, after the DM
event the star first evolves towards the ZAHB star as a He-rich
subdwarf. Then, when gravity has increased sufficiently, the star
crosses the wind limit causing homogeneous winds to halt, then diffusion
leads to an increase in the H abundance as the star evolves towards
the ZAHB (both processes occur on timescales of the order of 1
Myr). We note in particular (Fig.
\ref{fig:chem_t}) that  in only a few thousand years the H abundance rises from
its post-DM value to the values usually measured in He-sdO stars
(Str\"oer et al. 2007) and thus the correct H-abundances are
predicted. When the HB has been reached the evolutionary timescale
becomes more than a factor of 40 longer (65 to 90 Myr, see tab. 2) and
the star turns into a H-dominated sdB star. This possible evolutionary
channel is sketched in Fig. \ref{fig:channel}. In this case no
clustering of He-sdO stars should be expected close to the EHB, in
good agreement with the results of Str\"oer et al. (2007). It is
interesting to note that the effective temperatures expected for these
stars, when on the HB (and with a H-dominated atmosphere), agree with
those of the hotter sdB stars (which also have higher He abundances
than their cooler counterparts). The conversion of surface abundances
with C$>$N into surfaces with N$>$C as H diffuses outwards, might also
be related (see Figs. \ref{fig:chem_t} and  \ref{fig:channel}) to the
lack of C-rich He-sdO stars with lower $N_{\rm He}/N_{\rm H}$ values
in the study of Str\"oer et al. (2007). Within this scenario and in
the absence of any other channel that creates hot-subdwarfs, the
number ratio of He-sdO stars to hot ($T_{\rm eff}\gtrsim 33\,500$) sdB
stars should be of the order of 1/40 in a complete sample (although
higher ratios can still be possible if some process that disturbs
diffusion is present).  Once the star finishes the He-core burning
stage (or EHB), it moves above the wind limit again (see
Fig. \ref{fig:channel}) and homogeneous winds again become
possible. If this is so, since the H-rich layer is very thin, a new
conversion into a He-sdO star might happen (now at higher surface
gravities and temperatures). Finally, it is interesting to note in
Fig. \ref{fig:chem_t} that the C-abundance decrease rapidly by about a
factor of 25 and the star changes its surface composition from C$>$N
to C$<$N. This occurs while the star is still strongly H-deficient. We
note that, at this point (between t=1000 and 10000 yr in
Fig. \ref{fig:chem_t}), the star displays surface abundances
qualitatively similar (H$\sim 2.5\times 10^{-3}$, N$\sim 2\times
10^{-3}$ and C$<$N by mass fraction), although quantitatively
different, to those inferred by Lanz et al. (2004) for the peculiar
star LB 1766.

In the case of post-SM/SM$^*$ stars our simulation indicates that the
atmosphere should turn into a H-dominated one in about 1000 yr from
the moment in which homogeneous winds cease. Then, if homogeneous
winds do not exist at high gravities, post-SM/SM$^*$ stars would only be
He-rich while they are above the wind limit (see
Fig. \ref{fig:channel}). This is clearly not the case for most He-sdO
stars in the sample of Str\"oer et al. (2007) but it does describe the
observed surface properties of JL87 ---$T_{\rm eff}= 29\,000$ K and
log$g=5.5$, Lanz et al. (2004); $T_{\rm eff}= 26\,000$ K and log$g=4.8$,
Ahmad et al. (2007).

We emphasize that the possible conversion of He-sdO stars
into hot sdB stars in the HB discussed in the previous paragraphs
is  interesting, not only because it predicts the correct
distribution in the log $g-T_{\rm eff}$ diagram but also in view of
the significantly higher He abundances of sdB stars with $T_{\rm
eff}\gtrsim 33\,500$K compared with their cooler counterparts, which
might reflect differences in their previous evolution.

In contrast, if the effects of element diffusion were of minor
importance for all He-rich subdwarfs, then only the abundances
resulting from SM/SM$^*$ events would be consistent with the surface
properties measured for He-sdO stars (Str\"oer et al. 2007), unless DM
abundances differ strongly from those predicted by the standard MLT
theory (see Sect. 4.1).  We note that the fraction of stars undergoing
SM/SM$^*$ events corresponds to only a small fraction of hot-flashers
in standard sequences. In this context, our results in Sect. 4.3 indicate
that the role of chemical gradients should be analyzed to determine
whether the fraction of shallow-mixing episodes might be increased by
the effect of chemical gradients that tend to prevent the penetration
of convection into the H-rich envelope.

The hot-flasher scenario therefore can qualitatively predict the
correct surface properties and distribution of He-sdO stars. It is
then interesting to note that the lack of close binaries (with periods
of the order of days as observed for sdB stars) in He-sdO stars is
also expected naturally within the hot-flasher scenario. Very close
binaries with final periods of the order of days will indeed fill
their Roche lobes before reaching the RGB-tip and never undergo the
HeCF. Also, it is not clear if it would be possible to have
hot-flashers within a common envelope system. In any case, in order to
have a hot-flasher within a close binary system, a very fine tuning of
the initial masses and periods seems to be needed. Consequently, post
hot-flasher H-deficient stars should not be usual in these systems.
The lack of close binary systems with He-sdO components should then not
be taken as an argument against the hot-flasher and in favor of the
merger scenario. In contrast, wide systems would be a much better
way of distinguishing between both possible evolutionary origins for
He-sdO stars, because they should be more probable within the hot-flasher
picture.

\section{Final Remarks}
We have performed 1D stellar evolution simulations of the hot-flasher
scenario for a wide variety of cases and metallicities. Based on these
results, we have given a detailed description of the surface
characteristics of standard sequences. For these sequences, we have
studied how abundances could be altered by element diffusion
processes.  We have then extended the scope of our work by studying
deviations from standard assumptions. Our results can be summarized as
follows:

\begin{itemize}
\item We have carried out an ample exploration of the parameter
space of the hot-flasher scenario that will allow far more effective
tests of its predictions. Our results confirm the partial results by
Cassisi et al. (2003) and Lanz et al. (2004) for the deep-mixing and
shallow-mixing events respectively.

\item Our deep-mixing sequences have abundances similar to
those recently inferred for PG1544+488, confirming that this star can
be the result of a DM event. Our results also confirm the result of
Lanz et al. (2004) that JL87 abundances are compatible with a recent
SM event for this star. None of our sequences show abundances similar
to those inferred for LB 1766 and this object is therefore probably
not the result of a recent hot-flasher event. However, it is worth
noting that after a DM event, element diffusion might produce
abundances qualitatively similar to those of LB 1766.

\item Standard hot-flasher sequences predict C/N ratios in the range from
$\sim 9.5$ to $\sim 0.85$. This is particularly interesting in the
light of the N-rich and C-rich classes described by Str\"oer et
al. (2007), although no clear correlation is found between those
classes and our sequences in the $g-T_{\rm eff}$ diagram. In the
absence of weak winds, however, the lack of C-rich He-sdO stars with
lower He/H values in the sample of Str\"oer et al. (2007) can be
qualitatively understood in terms of  element diffusion.

\item The location of our sequences in the $g-T_{\rm eff}$ diagram show 
  qualitative agreement with the parameters inferred by Str\"oer et
  al. (2007) for He-sdO, while they predict (as already noted by
  Str\"oer et al. 2007) that these stars should cluster around the
  EHB, which is not observed.  An interesting (and plausible, in view
  of the results of Unglaub 2008) possibility for this discrepancy
  to be removed is if He-sdO stars evolve from a region in
  the $g-T_{\rm eff}$ diagram where homogeneous winds can exist, into a
  region where such winds are impossible.  This should lead to the
  onset of H-diffusion which, as we have showed in Sect 4.4, allows
  the conversion of He-sdO stars close to the ZAHB into H-dominated
  sdB stars even for the low abundances resulting from DM events.
  This scenario is extensively discussed in Sect. 5.

In the presence of weak homogeneous winds only the surface properties
arising from SM/SM$^*$ episodes would be consistent with those
inferred for He-sdO stars, unless DM abundances differ strongly from
those predicted by the standard MLT with diffusive convective mixing.

\item In all cases in which a conversion of He-sdO into hot H-rich 
sdB stars is possible, we have shown that the hot-flasher scenario
 reproduces correctly the observed properties (abundances and their
 distribution in the $T_{\rm eff}-g$ plane) of He-sdO stars and links
 them, as  immediate progenitors, to the hotter sdB stars.  This is
 interesting in view of the higher He-abundances of the hotter sdB
 stars.

\item We have argued that the lack of close binaries among He-sdO stars 
cannot be used as an argument against the hot-flasher scenario because this
dearth is also naturally expected within this scenario.

\item Our study of departures from standard assumptions shows that the role of
   chemical gradients as an extra barrier to the penetration of the
   inner convective zone into the H-rich envelope has to be studied in
   detail.  This is particularly important in determining if the
   fraction of SM/SM$^*$ episodes is higher than predicted by standard
   sequences. In contrast, extra-mixing processes alone does
   not seem to produce significant departures from standard
   predictions.
\end{itemize}

We conclude that the hot-flasher scenario is a viable explanation of
the formation of He-rich subdwarfs and that the understanding of some
key points needs to be clarified before a final statement can be
made. Special attention should be paid to analyzing the existence and
intensity of disturbing processes, such as winds and turbulence, in
the outer layers of He-rich subdwarfs. Also, the search for wide
binaries among He-sdO stars is necessary to distinguish between the
predictions of the merger and hot-flasher scenarios.  Finally, a
better understanding of the burning and mixing of H in the hot
interior (probably by means of hydrodynamical simulations of the
event) is badly needed.

\begin{acknowledgements}
 Part of this work was supported by PIP 6521 grant from CONICET. M3B thanks
 professor U. Heber for instructive correspondence, and professor O. Benvenuto
 and Dr. A. De Vito for useful comments on binary evolution. Also the
 Max-Planck Institut f\"ur Astrophysik in Garching and the European
 Association for Research in Astronomy are gratefully acknowledged for an
 EARA-EST fellowship under which this work was started.  Finally, we want
 to thank our referee, S. Cassisi, for his comments and suggestions which have
 improved the quality and readability of the article.
\end{acknowledgements}

{\sl Note Added in Proof:} After acceptance, the authors became aware
of the close binary nature of PG 1544+488 (Ahmad, A. \& Jeffery
C. S. 2007, ASPC, 391, 261). This fact raises serious doubts on a
possible deep-mixing origin for this star.

\end{document}